\documentclass[reprint, showkeys, onecolumn,
 amsmath,amssymb,
 aps
]{revtex4-2}

\usepackage{arydshln}
\usepackage{chngcntr}
\usepackage{hyperref}
\usepackage{url}
\usepackage{xcolor}
\usepackage{graphicx}
\usepackage{dcolumn}
\usepackage{bm}
\usepackage{amsfonts}
\usepackage{algorithmicx}
\usepackage{algorithm,algcompatible,lipsum}
\usepackage[english]{babel}
\usepackage[utf8]{inputenc}
\usepackage{algpseudocode}
\usepackage{arydshln}

\begin{document}

\title{Learning to swim in potential flow}

\author{Yusheng Jiao}

\affiliation{Department of Aerospace and Mechanical Engineering, University of Southern California, Los Angeles, CA 90089}
\author{Feng Ling}
\affiliation{Department of Aerospace and Mechanical Engineering, University of Southern California, Los Angeles, CA 90089}
\author{Sina Heydari}
\affiliation{Department of Aerospace and Mechanical Engineering, University of Southern California, Los Angeles, CA 90089}
\author{Nicolas Heess}
\affiliation{DeepMind, London}
\author{Josh Merel}
\affiliation{DeepMind, London}
\author{Eva Kanso}
\email{kanso@usc.edu}
\homepage{https://sites.usc.edu/kansolab/}
\affiliation{Department of Aerospace and Mechanical Engineering, University of Southern California, Los Angeles, CA 90089}

\begin{abstract}
Fish swim by undulating their bodies.  These propulsive motions require coordinated shape changes of a body that interacts with its fluid environment, but the specific shape coordination that leads to robust turning and swimming motions remains unclear. 
To address the problem of underwater motion planning, we propose a simple model of a three-link fish swimming in a potential flow environment and we use model-free reinforcement learning for shape control. We arrive at optimal shape changes for two swimming tasks: swimming in a desired direction and swimming towards a known target. This fish model belongs to a class of problems in geometric mechanics, known as \textit{driftless} dynamical systems, which allow us to analyze the swimming behavior in terms of geometric phases over the shape space of the fish. These geometric methods are less intuitive in the presence of \textit{drift}. Here, we use the shape space analysis as a tool for assessing, visualizing, and interpreting the control policies obtained via reinforcement learning in the absence of drift. We then examine the robustness of these policies to drift-related perturbations. Although the fish has no direct control over the drift itself, it learns to take advantage of the presence of moderate drift to reach its target.
\end{abstract}

\keywords{Fish swimming, Reinforcement learning, Sensorimotor control.}
\maketitle

\section{Introduction}

\begin{figure*}[!t]
\centering
\includegraphics[scale=1]{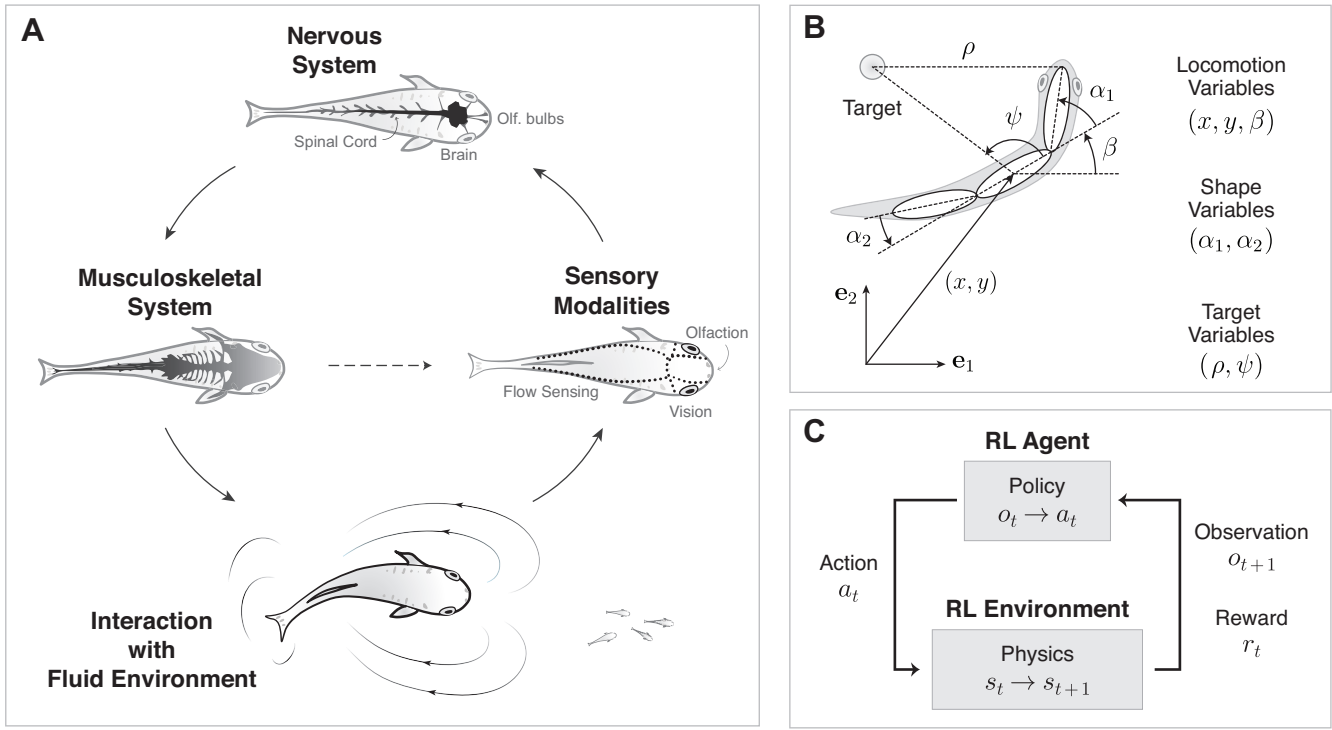}
\caption{\textbf{Model-free reinforcement learning and the three-link fish.} \textsf{A}. Illustration of sensorimotor feedback loops in fish. Motor commands generated in the nervous system activate the musculoskeletal system, resulting in deformations of the body. Body deformations, through interaction with the fluid environment, lead to swimming meanwhile sensory modalities provide sensory feedback to the nervous system. The dashed arrow between musculoskeletal and sensory systems indicates somatic sensing used to assess whether previous motor commands were successfully executed. Other reflexive or preflex signals could also be at play \cite{Dickinson2000,Engelmann2000,Tytell2011}. \textsf{B}. Three-link fish swimming in quiescent fluid. Locomotion variables $(x,y,\beta)$ are set in a lab fixed frame, while the shape variables $(\alpha_1,\alpha_2)$ and target variables $(\rho,\psi)$ are set in body frame symbolizing egocentric control and learning. \textsf{C}. To apply model-free reinforcement learning to our problem, we only need to set the appropriate state, observation, action, and reward variables based on our fish model.}
\label{fig:fishRL}
\end{figure*}

Fish swim through interactions of body deformations with the fluid environment.
A fish assimilates sensory information about its own body and the external environment and produces patterns of muscle activation that result in desired body deformations; see Fig.~\ref{fig:fishRL}A. 
How these sensorimotor decisions are enacted at the physiological level, at the level of neuronal circuits, remains unclear~\cite{Dickinson1999, Tytell2011, Merel2019a, Madhav2020}.
Animal models, such as the \textit{Danio rerio} zebrafish~\cite{Huang2013, Burgess2007, Privat2019}, as well as robotic and mathematical models~\cite{Marchese2014, Lauder2015}, provide valuable insights into the sensorimotor control underlying fish behavior.
Such understanding offers enticing paradigms for the design of artificial soft robotic systems in which the control is \textit{embodied} in the physics~\cite{Cianchetti2012, Laschi2012}. 
\textit{Embodied} systems sense and respond to their environment through a physical body~\cite{Pfeifer2006,Pfeifer2007}; physical interactions with the environment are thus vital for sensing and control.
In fish, the dynamics of the fluid environment is essential both at an evolutionary time scale -- in shaping body morphologies~\cite{van2013} and sensorimotor modalities -- and at a behavioral time scale. 
Fish bodies are tuned to exploit flows~\cite{Liao2003, Muller2003}. 
Body designs and undulatory motions have been examined in computational and semi-analytical models of fluid-structure interactions~\cite{Eloy2013, Kern2006, Mittal2006, Eldredge2006, Gazzola2015}, including models of body stiffness and neuromechanical control~\cite{Tytell2010,Hamlet2015}. The fish's ability to integrate multiple sensory modalities such as vision~\cite{Guthrie1986,Fernald1989,Douglas2012} and flow sensing~\cite{Engelmann2000, Ristroph2015} are essential to behaviors ranging from rheotaxis~\cite{Colvert2016,Colvert2017,Montgomery1997} to schooling~\cite{Filella2018,Li2020,Partridge1980,Katz2011}. 
Recent developments prove that machine learning techniques are highly effective in addressing problems of flow sensing and fish behavior~\cite{Gazzola2016,Gustavsson2017,Colabrese2017, Verma2018, Colvert2018, Alsalman2018, Weber2020, Gazzola2014}.

A central problem in fish behavior, which is also relevant for underwater robotic systems, is \textit{gait design} or \textit{motion planning}: what body deformations produce a desired swimming objective? The answer requires an understanding of how the numerous biomechanical degrees of freedom of the fish body are coordinated to achieve the common objective. Mathematically, this problem is often expressed in terms of an optimality principle: find control laws that optimize a desired objective, such as maximizing swimming speed or minimizing energetic cost~\cite{Kanso2005a, Kern2006, Eloy2013}.
But how these control laws are implemented in the nervous system, and how they are acquired via learning algorithms, are typically beyond the scope of such methods. As these optimization methods rely on an internal model of the dynamics, different computational results have been obtained by varying the specification of the physical model of the fish, the performance metric, and the control constraints~\cite{Kanso2005a, Kanso2005b, Kern2006, Eloy2013}. 

Model-free reinforcement learning (RL) of embodied systems offers an alternative framework for gait design that is mathematically and computationally tractable~\cite{Degris2012,Haith2013, Tsang2020}.
In this framework, the fish's world is divided into a body controlled by a \textit{learning agent} and an environment that encompasses everything outside of what the agent can control. 
The agent can be viewed as an abstract representation of the parts of the fish responsible for sensorimotor decisions. 
In RL, the agent must learn from experience in a trial-and-error fashion. 
Specifically, repeated interactions of the body with the environment enable the agent to collect \textit{sensory observations}, \textit{control actions}, and associated \textit{rewards}.
The goal of the agent is thus to learn to produce behavior that maximizes rewards, and the process is model-free when learning does not make use of either a priori or developed knowledge of the physics of the system. 
The learned feedback control law, called a \textit{policy}, is essentially a mapping from sensory observations to control actions. This mapping is nonlinear and stochastic, and, by construction, rather than providing a single optimal trajectory, it can be applied to any initial condition and transferred to conditions other than those seen during training such as when the body or fluid environment are perturbed~\cite{Sutton2018}. 

Here, we employ RL to design swimming gaits. We use an idealization in which the fish is modeled as an articulated body consisting of three links, with front and rear links free to rotate relative to the middle link via hinge joints~\cite{Kanso2005a,Kanso2005b,Eldredge2007,Melli2006,Nair2007,Hatton2010}.
In describing the physics of the fish, we cede the complexity of accounting for the full details of the fluid medium in favor of considering momentum exchange between the articulated body and the surrounding fluid in the context of a potential flow model~\cite{Kanso2005a,Kanso2005b,Melli2006,Nair2007}.
This model is a canonical example of a class of under-actuated control problems whose dynamics can be described over the actuation (shape) space using tools from geometric mechanics~\cite{Kanso2005a,Morgansen2001,Bloch1996,Murray1994,Kelly2012}. 
Specifically, swimming motions can be represented by the sum of a \textit{dynamic phase} or \textit{drift}, and a \textit{geometric phase} over the shape space of all body deformations \cite{Kanso2005b}. In the geometric phase case, a geometric \textit{connection}, defined as vector fields over the actuation shape space, maps infinitesimal shape changes into infinitesimal rigid motions of the whole body.
From a motion control perspective, this geometric framework is advantageous in that it provides tools for gait analysis and manual design of control policies over the full shape space~\cite{Hatton2010}. 
These geometric tools also provide an intuitive way for direct and interpretable visualizations of the RL-based policies. Here, we visualize the RL policies as vector fields over the shape space. This framework leads to a straightforward interpretation of the RL policy: given an observation of its own shape, the fish's action needs to follow the corresponding vector in the shape space.  A trajectory or realization of the RL policy arises from locally following these vectors to achieve a desired global task. These tools also allow us to probe the optimality and behavior of the RL policy in light of the physics of the system and in comparison to manually-designed policies over the fish shape space.  


\section{Mathematical Model of the Three-link Fish}
\label{sec:fish}

Consider a three-link fish as shown in Fig.~\ref{fig:fishRL}(B). Rotations of the front and rear links relative to the middle link are denoted by the angles $\alpha_{1}$ and $\alpha_{2}$ such that $(\alpha_1,\alpha_2)$ fully describe all possible body deformations. We constrain the swimming motions to a two-dimensional plane, and let $(x,y)$ and $\beta$ denote the net planar displacement and rotation of the fish body, such that $(\dot{x},\dot{y})$ and $\dot{\beta}$ represent the linear and rotational velocities of the fish in inertial frame (the dot notation $\dot{()} = \rm{d}()/\rm{d}t$ represents differentiation with respect to time $t$). We also introduce the linear velocity $(v_1,v_2)$ expressed in a co-rotating body frame attached to the center of the middle link,
\begin{equation}
\label{eq:vel}
v_1 = \dot{x} \cos\beta + \dot{y}\sin\beta, \qquad v_2 = -\dot{x} \sin\beta +\dot{y}\cos\beta.
\end{equation}
The total linear momentum $(p_x,p_y)$ and total angular momentum $\pi$ of the body-fluid system are expressed in the inertial frame, and they are related to their counterparts $(P_1,P_2)$ and $\Pi$ in body-fixed frame as follows,
\begin{equation}
\begin{split}
\label{eq:momenta_inertial}
p_{x} = P_{1} \cos\beta  -  P_{2} \sin\beta, \qquad
p_{y} =P_{1}  \sin\beta +  P_{2} \cos\beta, \qquad
\pi = \Pi.
\end{split}
\end{equation}
In potential flow, it is a known result that the total linear and angular momenta of the body-fluid system can be expressed in terms of the body geometry and velocity, via the so-called \textit{added mass} matrices~\cite{Lamb1945,Kanso2005a,Kanso2005b}. Expressions for the added mass matrices of the three-link fish are derived in detail in Appendix~\ref{app:fishmath}. The total momenta $(P_1,P_2)$ and $\Pi$ are given by
\begin{equation}
\label{eq:totalmomenta}
\left[\begin{array}{c}
P_{1} \\ P_{2} \\ \Pi
\end{array}\right] = \mathbb I_{\rm lock}
\left[\begin{array}{c}
v_1 \\ v_2 \\[1ex] \dot\beta
\end{array}\right] + \mathbb I_{\rm couple}
\left[\begin{array}{c}
\dot\alpha_{1} \\ \dot\alpha_{2} 
\end{array}\right],
\end{equation}
where $\mathbb I_{\rm lock}$ is the locked mass matrix at a given shape of the fish (see Eqs.~\ref{eq:Ilock}-\ref{eq:H}) and $\mathbb I_{\rm couple}$ is the mass matrix associated with shape deformations (see Eq.~\ref{eq:Icouple}). 

In the absence of external forces and moments on the fish-fluid system, the total momenta $(p_{x}, p_{y})$ and $\pi$ are conserved for all time.  Conservation of total momentum yields, upon inverting~\eqref{eq:momenta_inertial} and substituting into~\eqref{eq:totalmomenta},
\begin{equation}
\label{eq:eom}
\left[\begin{array}{c}
v_1 \\ v_2 \\[0.5ex] \dot\beta
\end{array}\right]  = 
\mathbb I^{-1}_{\rm lock}\left[\begin{array}{c c c}
\cos\beta & \sin\beta & 0\\
-\sin\beta & \cos\beta & 0\\
0 & 0 & 1
\end{array}\right]
\left[\begin{array}{c}
p_{x} \\ p_{y} \\ \pi
\end{array}\right] - 
\mathbb I^{-1}_{\rm lock}\mathbb I_{\rm couple}
\left[\begin{array}{c}
\dot\alpha_{1} \\ \dot\alpha_{2} 
\end{array}\right].
\end{equation}
If we further substitute~\eqref{eq:vel} into~\eqref{eq:eom}, we arrive at three coupled first-order equations of motion for $x$, $y$, $\beta$ given $\alpha_1$ and $\alpha_2$.  The control problem consists of finding the time evolution of shape changes $({\alpha}_1(t),{\alpha}_2(t))$ that achieve a desired locomotion task $(x(t),y(t))$ and $\beta(t)$. Specifically, a swimming gait is defined as a cyclic shape change $({\alpha}_1(t),{\alpha}_2(t))$, with period $T$, that results in a net swimming  $(x(t),y(t))$ or turning $\beta(t)$  of the fish body. 

This model is a canonical example of a class of under-actuated control problems whose dynamics can be described over the \textit{shape space}  using tools from geometric mechanics~\cite{Kanso2005a,Morgansen2001,Bloch1996,Murray1994}. 
On the right-hand side of~\eqref{eq:eom}, the first term represents a \textit{dynamic phase} or \textit{drift} and the second term represents a \textit{geometric phase} over the fish shape space $(\alpha_1,\alpha_2)$ \cite{Kanso2005b}.
The geometric phase is best described in terms of the \textit{local connection} matrix $\mathbb A$~\cite{Kanso2005b,Hatton2010}, which is a function  only of the shape variables $\alpha_1$ and $\alpha_2$,
\begin{equation}
\label{eq:connection_matrix}
\mathbb A =
\left[
\begin{array}{c c}
A_{11} & A_{12}\\
A_{21} & A_{22}\\
A_{\beta 1} & A_{\beta 2}
\end{array}
\right] := -\mathbb I_{\rm lock}^{-1}\mathbb I_{\rm couple}.\\
\end{equation}
Each row of $\mathbb A$ describes a nonlinear vector field over the shape space, giving rise to three vector fields $\mathbf{A}_1 \equiv (A_{11}, A_{12})$, $\mathbf{A}_2 \equiv (A_{21}, A_{22})$, and $\mathbf{A}_\beta \equiv (A_{\beta 1}, A_{\beta 2})$ over the   $(\alpha_1,\alpha_2)$ plane  as shown in Fig.~\ref{fig:colormap}(A). In driftless systems, net locomotion is fully controlled by the fish shape changes as dictated by the connection matrix $\mathbb{A}$. However, in the presence of drift, shape control is not sufficient to determine the fish motion in the physical space, which is then affected by the drift term in~\eqref{eq:eom}.

\section{Geometric phases}
\label{sec:Geometry}
\begin{figure*}[t]
\centering
\includegraphics[scale=1]{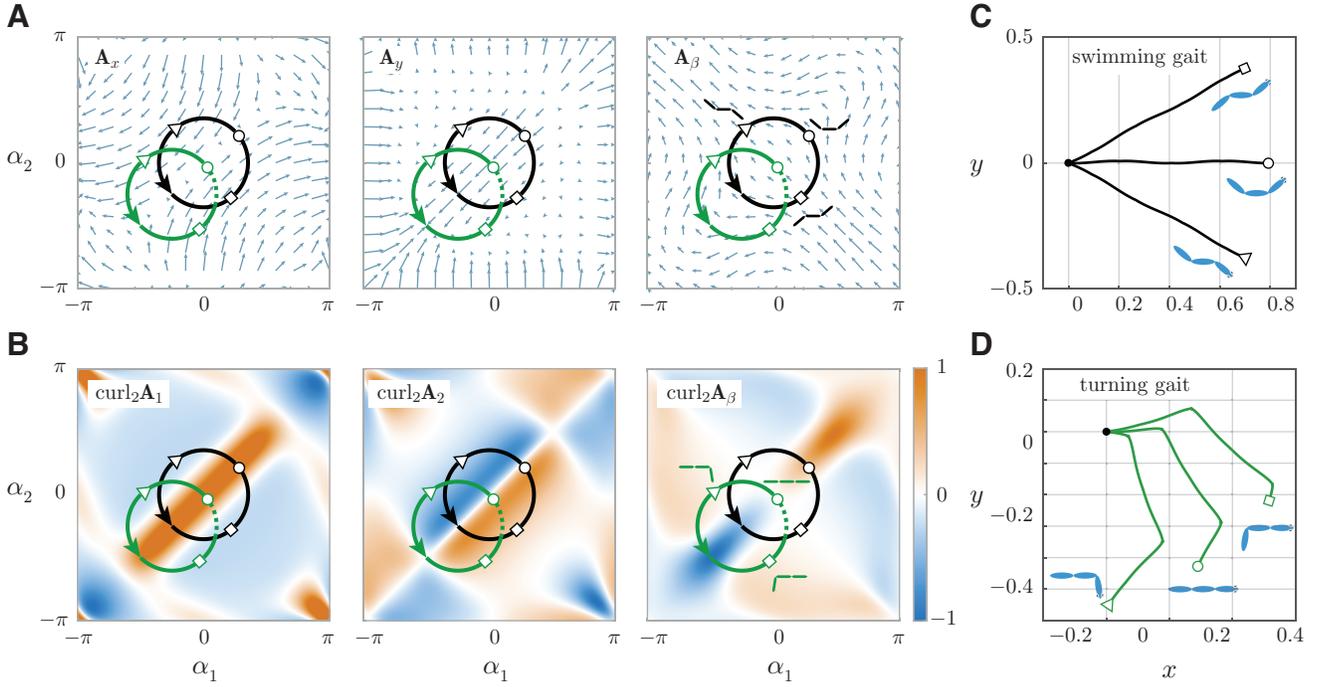}
\caption{\textbf{Using connection matrix for simple gait design.} \textsf{A} and \textsf{B}. Rows of connection matrix $\mathbb{A}$ give us three vector fields, $\mathbf{A}_1\equiv{\partial v_1}/{\partial \alpha_i}$, $\mathbf{A}_2\equiv{\partial v_2}/{\partial \alpha_i}$, and $\mathbf{A}_\beta\equiv{\partial \dot\beta}/{\partial \alpha_i}$, the curl of which yield three corresponding scalar fields. Magnitude of the scalar fields are normalized to be within $[-1,1]$. Value of the scalar fields can facilitate the design of simple swimming and turning gaits, as shown above by black and green circles respectively. Six body configurations, each corresponding to points marked by black and green {\large $\circ$}, {\scriptsize $\triangle$, $\Box$}, are sketched for additional clarity. \textsf{C} and \textsf{D}. Fish that start with body centered at the origin of the x-y plane and follow the same gait circle swim/turn in different directions in the physical space depending on initial body shapes; the initial shape is depicted in blue at the end of each trajectory.}
\label{fig:colormap}
\end{figure*}
Geometric phases are defined as the net locomotion $(x,y,\beta)$ that results from prescribed cyclic shape changes in the $(\alpha_1,\alpha_2)$ plane at zero total momentum (no drift).  
Inverting~\eqref{eq:vel} and substituting~\eqref{eq:connection_matrix} into~\eqref{eq:eom} at zero total momentum, we arrive at
\begin{equation}
\label{eq:eom_zeromom}
\begin{split}
\left[\begin{array}{c c c}
\cos\beta & \sin\beta & 0\\
-\sin\beta & \cos\beta & 0\\
0 & 0 & 1
\end{array}\right]
\left[\begin{array}{c}
\dot{x} \\ \dot{y} \\[0.5ex] \dot\beta
\end{array}\right]  =  \left[
\begin{array}{c c}
A_{11} & A_{12}\\
A_{21} & A_{22}\\
A_{\beta 1} & A_{\beta 2}
\end{array}
\right]
\left[\begin{array}{c}
\dot\alpha_{1} \\ \dot\alpha_{2} 
\end{array}\right] .
\end{split}
\end{equation}
Motions $(x,y,\beta)$ in the physical space are obtained by integrating~\eqref{eq:eom_zeromom} with respect to time. Rotations are directly proportional to the line integral of the vector field $\mathbf{A}_\beta$ as evident by integrating the last equation in~\eqref{eq:eom_zeromom},
\begin{equation}
\label{eq:beta}
\beta(T) - \beta(0)  = \oint_C\mathrm d\beta = \oint_C \left(A_{\beta 1}\mathrm d\alpha_1 + A_{\beta 2}\mathrm d\alpha_2\right).
\end{equation}
Here, $T$ is the time-period for going around the closed trajectory in the shape space once. Using Green's Theorem, we get
\begin{equation}
\begin{split}
\beta(T) - \beta(0)   = \iint \left(\frac{\partial A_{\beta 2}}{\partial \alpha_{1}}- \frac{\partial A_{\beta 1}}{\partial \alpha_{2}}\right) \mathrm d \alpha_1\mathrm d \alpha_2 = \iint {\rm curl}_{\rm 2} (\mathbf{A}_\beta) \mathrm d \alpha_1\mathrm d \alpha_2 .
\end{split}
\end{equation}
Here, ${\rm curl}_{\rm 2}$ denotes the two-dimensional curl as a scalar field over the $(\alpha_1,\alpha_2)$ plane. The scalar field $\textrm{curl}_2 (\mathbf{A}_\beta)$ provides an intuitive tool for understanding the effect of a cyclic shape deformation on the net rotation of the fish: net rotations are proportional to the integral of $\textrm{curl}_2 (\mathbf{A}_\beta)$ over the area enclosed by a closed shape trajectory; see Fig.~\ref{fig:colormap}. 
However, translational motions $(x,y)$ are not directly proportional to the area integrals of $\textrm{curl}_2(\mathbf{A}_1)$ and $\textrm{curl}_2(\mathbf{A}_2)$, but to a combination of all three integrals coupled  through the fish rotational dynamics as evident from~\eqref{eq:eom_zeromom}. This coupling is clear when the equations of motion for $x$ and $y$  in \eqref{eq:eom_zeromom} are written in scalar form,
\begin{equation}
\label{eq:eom_zeromom_trans}
\begin{split}
\dot{x} & = \left[ A_{11}\dot{\alpha}_1 + A_{12}\dot{\alpha}_2 \right]\cos\beta -  \left[ A_{21}\dot{\alpha}_1 + A_{22}\dot{\alpha}_2 \right]\sin\beta, \\
\dot{y} & = \left[ A_{11}\dot{\alpha}_1 + A_{12}\dot{\alpha}_2 \right]\sin\beta + \left[ A_{21}\dot{\alpha}_1 + A_{22}\dot{\alpha}_2 \right]\cos\beta.
\end{split}
\end{equation}
Despite this complication, the scalar fields defined by $\textrm{curl}_2(\mathbf{A}_1)$, $\textrm{curl}_2(\mathbf{A}_2)$,
and $\textrm{curl}_2(\mathbf{A}_\beta)$, shown in Fig.~\ref{fig:colormap}(B), are informative of the net translational $(x,y)$ and rotational $\beta$ motions of the fish. 
To illustrate the utility of these curl fields, we show two examples of cyclic shape changes depicted in black and green lines. A fish changing its shape following the black line undergoes zero net rotation because the area integral of $\textrm{curl}_2(\mathbf{A}_\beta)$ is identically zero, but it  swims forward in the $(x,y)$ plane. The net displacement per period is a conserved quantity, whereas the direction of motion depends on a combination of the fish initial shape $(\alpha_1(0),\alpha_2(0))$ and initial orientation $\beta(0)$ as shown in Fig.~\ref{fig:colormap}(C). Here, we consider three different initial shapes, depicted by the markers $\large\circ$, {\scriptsize $\triangle$, $\Box$}, all at $\beta(0) = 0$. Similarly, shape deformations following the green line lead to net reorientations in the physical space, as shown in Fig.~\ref{fig:colormap}(D).
Evidently, no net motion occurs if the shape trajectory is degenerate, that is, does not enclose an area in the shape space. Further a re-scaling of time does not affect the net motion of the fish, only the speed at which the fish completes these cyclic shape changes.

In Fig.~\ref{fig:colormap} and hereafter,  the equations of motion are non-dimensionalized using the total length of the fish and the total mass in the head-to-tail direction of the straight fish as the characteristic length and mass scale.  Specifically, we set the total mass of the fish body to be equal to the added mass (actual mass of the  fish is negligible). We leave the time scale unchanged because the characteristic time depends on the speed of shape changes, which is a control parameter to be determined by the controller.

The scalar fields curl$_2 (\mathbf{A}_1)$, curl$_2(\mathbf{A}_2)$, curl$_2(\mathbf{A}_\beta)$ over the shape space encode information about the net locomotion of the fish in a driftless environment, and can be used to design simple swimming and turning gaits as shown in Fig.~\ref{fig:colormap}. However, these geometric tools do not allow for a straightforward design of control policies for arbitrary motion planning~\cite{Melli2006,Hatton2010}, and they are even less instructive in the presence of drift. Next we consider an RL driven approach for motion control.


\section{Motion control via reinforcement learning}
\label{sec:RL}

We use RL to train the three-link fish on two different tasks: (i) to swim parallel to a desired direction in a driftless environment; (ii) to swim towards a target point located at a distance $\rho$ and angular position $\psi$ from the fish nose, with $\rho$ and $\psi$ expressed in the fish frame of reference (Fig.~\ref{fig:fishRL}B). Given the rotational symmetry of the fish-fluid space, 
in the first task, we fix the desired direction to be parallel to the $x$-axis without loss of generality.
For the second task, we first train the fish in a driftless environment, then introduce drift and train again in the presence of drift. 
The first task allows for direct comparison of the performance of the trained policy to manually-designed policies in the context of geometric mechanics as described in \S~\ref{sec:Geometry}. The second task allows for evaluation and comparison of the performance of the driftless and drift-aware policies under environmental perturbations.

Central to any RL implementation are the notions of the \textit{state} of the system, the \textit{observations} given to the learning agent, the \textit{actions} taken by the agent, and the \textit{rewards} given to the agent in light of its behavior. The state $s_t$ of the fish-fluid-target system at a time $t$ is given by the fish position and orientation in inertial frame $(x,y,\beta)$, its shape $(\alpha_1,\alpha_2)$, and the target position relative to the fish $(\rho,\psi)$; see Fig.~\ref{fig:fishRL}(B,C). As sensory input, we provide the fish a set of observations based on its proprioception of its own shape $\alpha_1$ and $\alpha_2$, as well as an egocentric observation of the task, namely for controlling the direction of swimming, the fish knows the desired swimming direction relative to itself $\psi = - \beta$, and for swimming towards a target, it knows the relative angular position $\psi$ of the target point. This yields a set of observations $o_t=(\alpha_1,\alpha_2,\psi)_t$. 
Additionally, when training in the presence of drift, the magnitude and direction of the drift vector $(p_x,p_y)$ are also provided as observations.
As control action, the fish has direct control of its shape using the rate of shape changes  as actions $a_t = (\dot\alpha_1,\dot\alpha_2)_t$. With this choice of action, the control can be projected onto the shape space and directly compared to the geometric mechanics approach. We constrain the value of the actions to be between $-1$ and $1$ rad per unit time, and we impose limits on the joint-angle so that $\alpha_{1}$ and $\alpha_{2}$ are only allowed to change between $-2\pi/3$ and $2\pi/3$ rad to avoid self-intersection.

In RL, the decision making process is modeled as a stochastic control policy $\pi_\theta(a_t|o_t)$ that produces actions $a_t$ given observations $o_t$ of the state of the fish-fluid system. The policy is parameterized by a set of parameters $\theta$ to be optimized. An optimal policy is learned to produce behavior that maximizes rewards. We use a dense shaping reward, that is, the fish is given a reward at every decision time step. Specifically, we set the reward to be the distance the fish travels in the desired direction towards the target state. For learning to swim parallel to the $x$-axis, we use the reward $r_{t} = x_{\rm{nose},t+1} - x_{\rm{nose},t}$, which is the change in the fish nose position along the $x$-axis. For learning to swim towards a target, the reward $r_t = \rho_t - \rho_{t+1}$ is based on the change in the relative distance $\rho$ from the fish nose to the target. The \textit{return} $R_t = \sum_{{t}'}^\infty \gamma^{t'-t} r_{t'}$ is defined as the infinite horizon objective based on the sum of discounted future rewards, where $\gamma \in [0, 1]$ is known as the discount factor; it determines the preference for immediate over future rewards. We set $\gamma = 0.99$ to make the fish foresighted.
The goal is to arrive at an optimal set of parameters $\theta$ that maximizes the \emph{expected} return $\mathcal{J}(\pi_\theta) = \mathbb{E}_{\pi}\big[\sum_{t=0}^{\infty} \gamma^{t} r_t\big]$ for a distribution of initial states. Here, the expectation is taken with respect to the distribution over trajectories $\pi(\tau)$ induced jointly by the fish dynamics, viewed as a partially-observable Markov decision process, and the policy $\pi_\theta(a_t | o_t)$. One approach to solving this optimization problem is to use a policy gradient method that computes an estimate of the gradient $\nabla_\theta \mathcal{J}$ for learning. Policy gradient methods are widely used to learn complex control tasks and are often regarded as the most effective reinforcement learning techniques, especially for robotics applications~\cite{Williams1992, Sutton2000, Kakade2002a, Peters2008, Peters2006}. Here, we use a specific class of policy gradient methods, known as \textit{actor-critic} methods~\cite{Konda2000, Haarnoja2018} where the fish learns simultaneously a policy (actor) and a value function (critic). We implement this method using the clipped advantage Proximal Policy Optimization (PPO) algorithm proposed in \cite{Schulman2017b}. This algorithm ensures fast learning and robust performance by limiting the amount of change allowed for the policy within one update. A pseudo-code implementation of the PPO algorithm and additional implementation details are provided in Appendix~\ref{app:ppo}.


\section{Training the Fish to Swim}
We trained the three-link fish to (i) swim parallel to the $x$-direction in a driftless environment; (ii) swim towards a target point  in the absence and presence of drift. 
We refer to the first task as \textit{direction-control} for short, and the second task as \textit{naive} and \textit{drift-aware} \textit{target-seeking}, respectively, based on their awareness of drift.

For the purpose of efficient training we imposed a finite time interval, following which the system state was reset to the initial state for a new round of training. Each round is referred to as an \textit{episode}. In all training episodes, we initialized the fish center to be at the origin of the inertial frame, and we initialized the shape angles $\alpha_1, \alpha_2$ and body orientation $\beta$ by sampling from a uniform distribution over all permissible angles to maximize the chances for robust learning. We fixed the maximum episode length to 150 time steps, with no early termination allowed. In training the target-seeking policies, the target was initially placed at a fixed distance (three-unit length) from the fish center but at a random orientation. For the drift-aware policy, a drift was introduced in the form of a non-zero total linear momenta $p_{x}, p_{y}$ in \eqref{eq:eom}. The drift magnitude was sampled from a uniform distribution between 0 and 0.15 and its direction from a uniform distribution between 0 and 2$\pi$, such that the drift was kept constant within a training episode.
Based on the non-dimensional scales introduced in section \ref{sec:Geometry}, one unit of drift aligned parallel to a straight fish causes it to translate one-unit length per unit-time.

For the direction-control policy, we performed 24 runs of RL training with 10,000 episodes in each run. 
The training process is illustrated in terms of reward evolution, calculated by taking the sum of all rewards in an episode, in Appendix B, Fig.~\ref{fig:rewVsEpi}(A). 
There was some variability among the seeds, but most trained policies performed very well; only a single policy did not converge by the end of training episodes. 
Note that fluctuations in reward after policy convergence are partly due to the stochasticity built into the policy itself and partly due to variation in task difficulty given the random initial conditions: different initial conditions require different amount of time and effort for the fish body to align with the $x$-axis.

It is worth pointing out here the training results of the direction-control policy were affected significantly by the episode length. In order to swim in a desired direction starting from an arbitrary initial orientation, the fish has to first turn in that direction, then swim forward. Policies trained with longer episodes performed better in the swimming portion of the trajectory but failed to make large-angle turns, as training data collected on swimming significantly outweighed those collected on turning. On the other hand, policies trained with shorter episodes made turns of any angle, but were less likely to swim straight after turning. We chose the episode length, following several trial and error trainings, to be 150 as a reasonable compromise to learn to both turn and swim effectively.

The evolution of rewards during training of the two target-seeking policies are plotted in Fig.~\ref{fig:rewVsEpi}(B), each with 20 runs of the RL algorithm and 15,000 episodes in each run. The naive policy converged faster than the drift-aware policy, and both policies converged slower than the direction-control policy. These results indicate that the task itself, as well as variations in the environment and number of observations affect the convergence rate, that is, the learning difficulty. Note that the numerical value of the reward is not directly comparable between policies for different tasks.

We evaluated the performance of the trained policies by testing them under two type of conditions: conditions similar to those seen during training and perturbed conditions not seen during training, as discussed next.

\section{Behavior of Trained Fish}
\label{sec:fishRLeval}

\begin{figure*}[!t]
\centering
\includegraphics[scale=1]{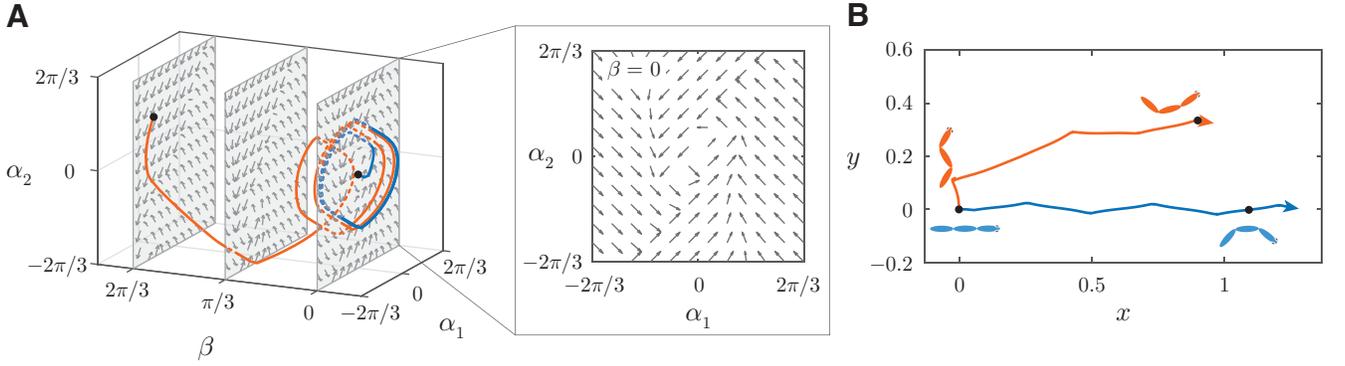} 
\caption{\textbf{Visualizing the direction-control RL policy.} \textsf{A}. Given the direction-control task, we visualize the mean RL policy actions $(\dot\alpha_1,\dot\alpha_2)$ as vector fields in the observation space of $(\alpha_1,\alpha_2,\beta)$. Two example observation trajectories starting at $\beta(0)=0,\alpha_{1}(0) = 0,\alpha_{2}(0)=0$ (blue) and $\beta(0)=2\pi/3,\alpha_{1}(0) = -\pi/3,\alpha_{2}(0)=\pi/3$ (orange) are plotted with slices of the mean policy at $\beta = 0, \pi/3, 2\pi/3$. The inset shows a flattened view of the slice at $\beta=0$. \textsf{B}. Physical space trajectories of the corresponding examples shown in \textsf{A} are plotted together with the fish body configurations at the starting point and chosen points near the ends. }
\label{fig:direction}
\end{figure*}
\begin{figure*}[!t]
\centering
\includegraphics[scale=1]{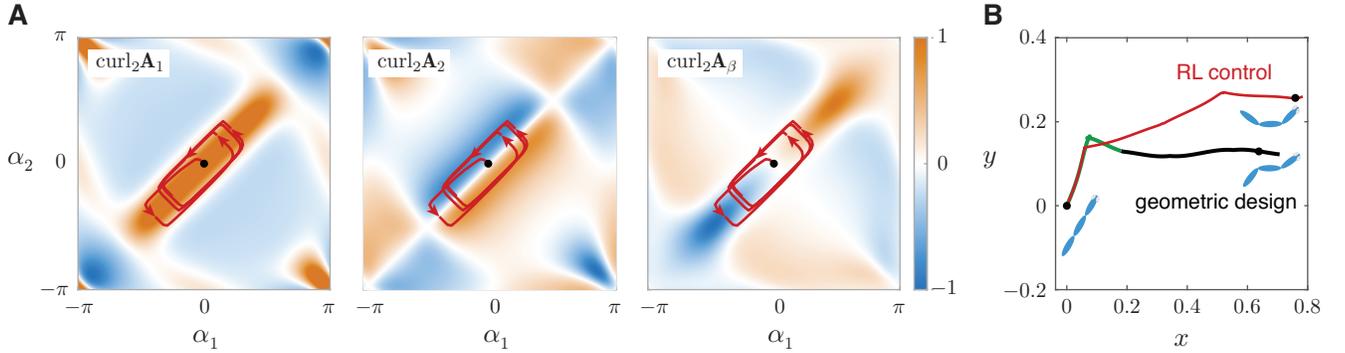} 
\caption{\textbf{RL provides smooth transition between turning and swimming gaits.} 
\textsf{A}. The shape space trajectory of the fish reorienting itself to swim parallel to the $x$-axis produced by mean RL policy is superimposed to the scalar curl fields from Fig.~\ref{fig:colormap}(B). Note that this trajectory starts off-centered and smoothly moves to cycles that are symmetric about the origin. \textsf{B}. The physical space trajectory due to the mean RL policy (red) in comparison to a manually patched turning-to-swimming trajectory (green and black) using circular gaits in Fig.~\ref{fig:colormap}. Without further fine-tuning on the shape of the gaits, the manually patched result shows a more abrupt and unnecessary turn angle. In both simulations, the fish start in a straight configuration with an orientation of $\beta(0) = \pi/3$ at the origin. }
\label{fig:gaitsCompare}
\end{figure*}

To visualize the RL direction-control policy, we plot in Fig.~\ref{fig:direction}(A) the action vector fields $(\dot\alpha_1,\dot\alpha_2)$ over the observations space $(\alpha_1,\alpha_2,\beta)$. These vector fields depend on the orientation $\beta$ of the fish in the physical space such that the control policy $(\dot\alpha_{1},\dot\alpha_{2})$ forms a \textit{foliation} over $\beta$. Three slices of this foliation are highlighted. The right hand side of Fig.~\ref{fig:direction}(A) provides a closer look at the policy slice at $\beta=0$; the arrows indicate the mean actions advised by the policy for a given set of observations $\alpha_1,\alpha_2$ at $\beta=0$. 
In Fig.~\ref{fig:direction}(B), we show the details of two trajectories in the physical space starting from two distinct configurations. In the first test, the fish starts at zero orientation, $\beta(0) = 0$, in a straight shape, $(\alpha_{1}(0),\alpha_{2}(0)) = (0,0)$. The goal of the fish is simply to swim forward. In the second test, the initial orientation and shape are $\beta(0) = 2\pi/3$ and $(\alpha_{1}(0),\alpha_{2}(0)) = (-\pi/3,\pi/3)$, from which the fish needed to turn and swim along the $x$-axis. 
In both cases, the fish is able to turn to the desired direction and swim steadily. 
%
In Fig.~\ref{fig:direction}(A), we highlight the corresponding trajectories in the $(\alpha_1,\alpha_2,\beta)$ space.  As the fish moves through the physical space, $\beta$ changes causing the fish to take actions from distinct slices of the foliation of action vector fields. 
Both trajectories tend to the same periodic swimming cycle around $\beta=0$, indicating that the control actions are similar once both fish are aligned with the desired swimming direction.

\begin{figure*}[t!]
\centering
\includegraphics[scale=1]{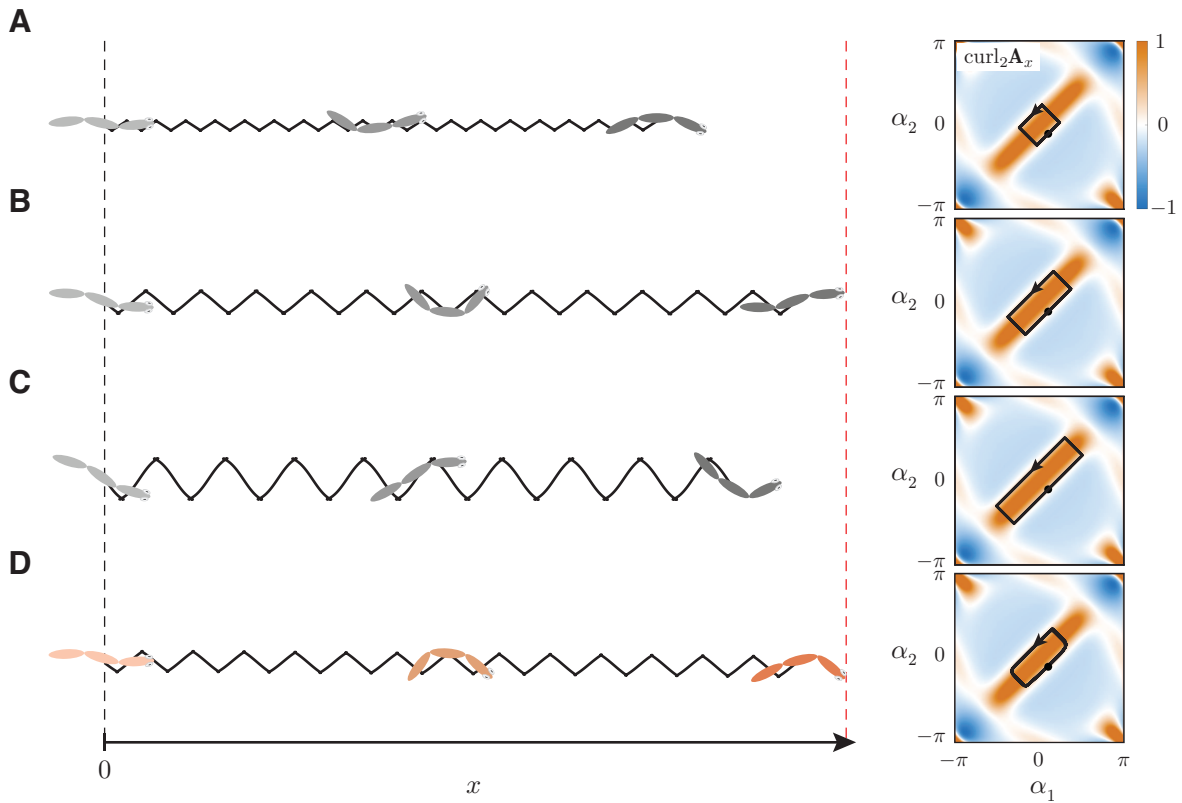} 
\caption{\textbf{Racing against the RL fish.} To test the optimality of our direction-control RL policy, we compare forward swimming performance between the geometrically-designed gaits and RL results. In \textsf{A-C} we use rectangular gaits with length equal to 1.2 (small), 2.4 (medium), and 3.6 (large), respectively, and width all equivalent to shape space trajectory following the mean actions from RL policy in \textsf{D}. All shape space trajectories are shown superimposed on top of the $\mathrm{curl}_2\mathbf{A}_1$ field on the right. Physical space trajectories on the left show that the mean RL policy achieves its excellent performance by choosing an optimal amount of lateral oscillation during forward swimming, while the small and large rectangular gaits move slower due to either insufficient or overwhelming side-way motion. Note that fish in \textsf{A-C} are initialized with the same shape but slightly different initial orientations to ensure they all swim in exactly the $x$-direction. In addition, all fish utilize the maximum actions allowed at each time step. }
\label{fig:fishRace}
\end{figure*}
\begin{figure*}[!t]
\centering
\includegraphics[scale=1]{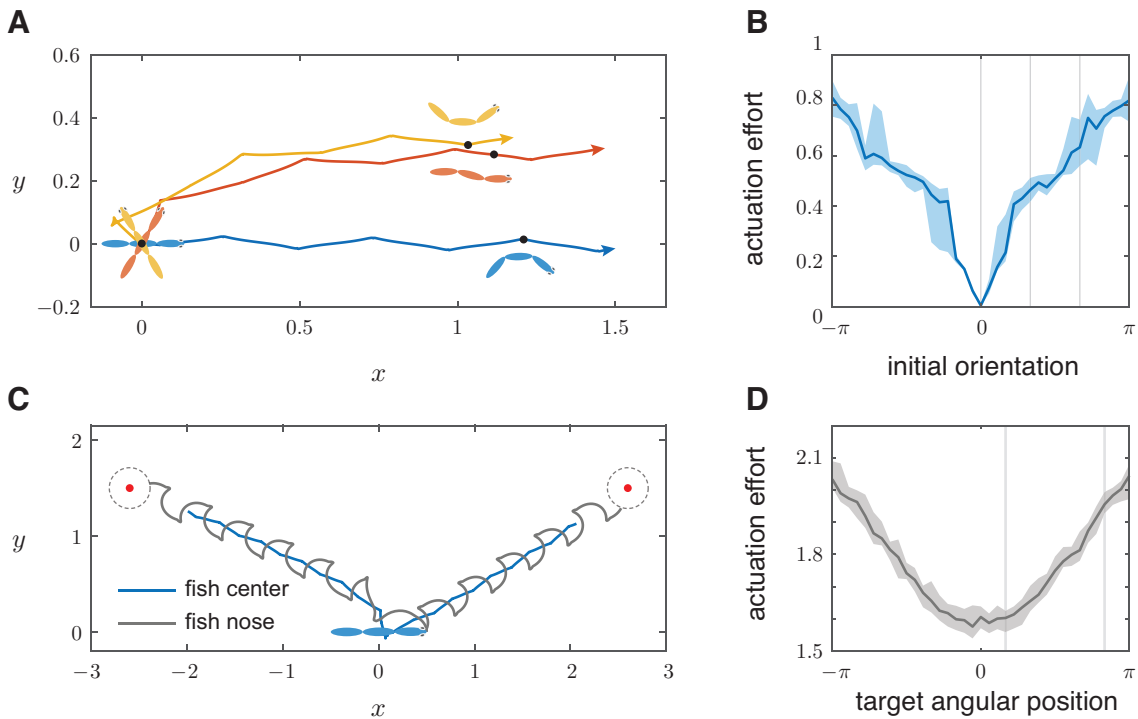} 
\caption{\textbf{Performance of RL policies in a driftless environment.} \textsf{A}. Swimming trajectories (center of mass) arrived at by the mean action of the direction-control policy are shown for fish starting at orientations $\beta = 0$ (blue), $\pi/3$ (red), $2\pi/3$ (yellow). \textsf{B}. The actuation effort of reorientation is roughly proportional to the absolute value of the initial fish orientation due to the amount of turning maneuvers required. Results shown are based on 25 stochastic policy roll-outs per tested orientation angle. \textsf{C}. Center of mass (blue) and nose (grey) swimming trajectories using the mean action of the naive target-seeking policy are shown with two targets located at an angular position of $\pi/6$ and $5\pi/6$. The fish is considered to have reached the target when its nose is within $\epsilon=0.2$ distance from the target (dotted circles). \textsf{D}. Actuation effort needed to perform shape increases as the target angular position changes away from $0$. Results shown use 25 stochastic policy roll-outs per tested target angle. Note that solid lines and shaded regions of \textsf{B} and \textsf{D} show the median results and variations between 25 to 75 percentile, respectively. In \textsf{B}, performance is calculated based on the actuation effort it takes for the fish to turn in the $x$-direction, while in \textsf{D} it is based on the effort it takes to reach the target.}
\label{fig:driftless}
\end{figure*}
\begin{figure*}[t!]
\centering
\includegraphics[scale=1]{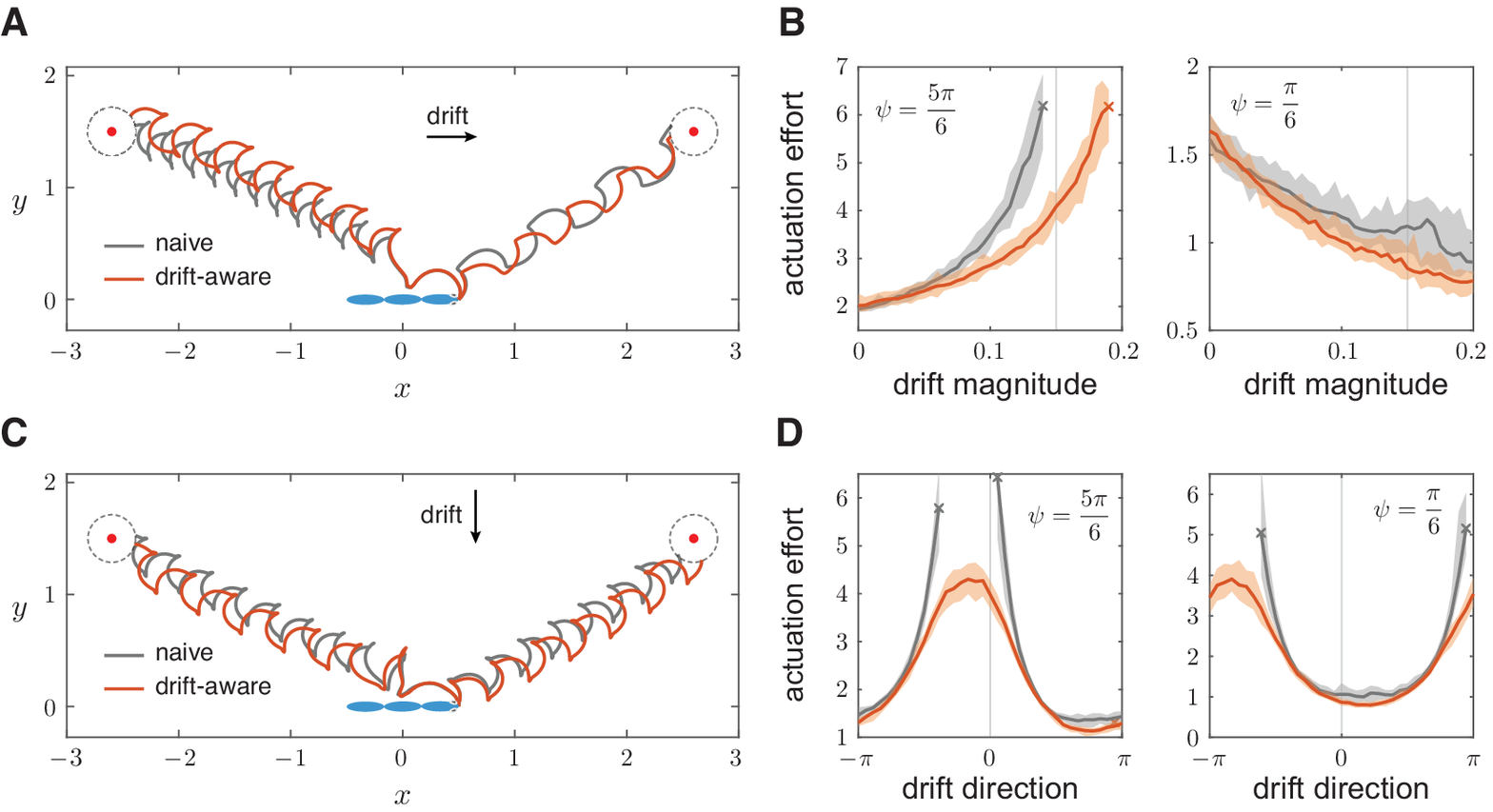}
\caption{\textbf{Performance of target-seeking policies in the presence of drift.} We compare the swimming trajectories and actuation efforts of target-seeking policies for different drift magnitudes and direction. \textsf{A} and \textsf{C}. Naive policy (grey) and drift-aware policy (orange) produce similar average actions in environments with constant (0.1) drift in the positive $x$-direction and negative $y$-direction, respectively. Both panels showcase trajectories reaching targets located three unit-lengths away with angular position $\psi$ of $\pi/6$ and $5\pi/6$. \textsf{B}. For a fixed drift direction (positive $x$-direction), actuation effort as a function of drift magnitude evolves differently depending on target angular position $\psi$. When drift is against the direction of the target (left), both policies fail to reach the target on average for large enough drift (cross marks). No failures are observed when drift facilitates swimming towards the target (right). In both cases, the drift-aware policy significantly outperforms the naive policy at large drift magnitudes by saving actuation effort. Intriguingly, inclusion of extra observations in drift-aware policies seems to result in slightly suboptimal performance when the drift magnitude is very small. \textsf{D}. With the drift magnitude fixed to $0.15$, the naive policy fails to reach the target if the drift direction is near the opposite end of the target angular position $\psi$. However, at this drift magnitude, the drift-aware policy can still reach both targets regardless of the drift direction. Note that solid lines and shaded regions of \textsf{B} and \textsf{D} show the median results and variations between 25 to 75 percentile, respectively. }
\label{fig:target}
\end{figure*}

We further explore the shape changes undertaken by the second (red) fish by superimposing these shape changes onto the scalar fields 
scalar fields curl$_2\mathbf A_{1}$, curl$_2\mathbf A_{2}$ and curl$_2\mathbf A_{\beta}$ introduced in Section~\ref{sec:Geometry}; see Fig.~\ref{fig:gaitsCompare}(A).
The corresponding motion in the physical space is depicted in red in Fig.~\ref{fig:gaitsCompare}(B).
The shape deformations produced by the RL policy can be interpreted as consisting of two regimes: an initial turning regime followed by a forward swimming regime. The turning regime is indicated by the initial portion of the shape trajectory enclosing most of the blue area in the curl$_2\mathbf A_{\beta}$ image; the integral in~\eqref{eq:beta} along this portion of the trajectory is negative, leading to a clockwise rotation. The swimming regime is indicated by the periodic shape changes enclosing the rectangular orange portion in the curl$_2\mathbf A_{1}$ image. The area integral of curl$_2\mathbf A_{1}$  along this portion of the trajectory is positive, whereas the corresponding area integrals of curl$_2\mathbf A_{2}$ and curl$_2\mathbf A_{\beta}$ are identically zero, leading to net motion in the positive $x$-direction. These shape deformations and resulting turning and swimming motions can be compared to a manually-designed shape trajectory based on the turning (green) and swimming (black) gaits in Fig.~\ref{fig:colormap}(A,B). Specifically, starting from a straight fish configuration, we follow the solid portion of the green trajectory (turning) in Fig.~\ref{fig:colormap}(A,B), and transition into the black trajectory (swimming) at the second intersection of the green and black shape trajectories. The resulting motion in the physical space is superimposed onto Fig.~\ref{fig:gaitsCompare}(B). 
Both the RL and manually patched gaits lead the fish  to turn and swim parallel to the $x$-axis, however, 
in the RL produced motion,  the transition between turning and swimming smoother.
Note that, in the swimming regime, the RL policy produces cyclic shape deformations that do not maximize the area integral of curl$_2\mathbf{A}_1$ (shape trajectory does not enclose the whole orange portion in the curl$_2\mathbf{A}_1$ image). 
Indeed, maximizing this integral is not optimal for forward swimming as discussed next. The key lies in the fact that the physics of the problem, specifically, the rotational motion of the fish, couples displacements in the $x$- and $y$- direction. Our model-free RL policy captures this effect with no explicit knowledge of the physics of the system.

To better explain this optimality of the RL policy, we manually prescribed cyclic shape changes that follow rectangular trajectories reminiscent of the trajectory generated by the RL policy for forward swimming. 
The manually-designed shape trajectories all share the same width as the RL policy albeit at different lengths, namely, 1.2 (small), 2.4 (medium), and 3.6 (large) to enclose increasingly larger regions of the orange portion of the curl$_2\mathbf{A}_1$; see right column of Fig.~\ref{fig:fishRace}. 
These cyclic deformations result in net displacements in the $x$-direction with zero-mean excursions in the $y$-direction. The $y$-excursions are due to the fact that, even though the area integrals of curl$_2\mathbf{A}_2$ and curl$_2\mathbf{A}_\beta$ over the regions enclosed by these shape trajectories are identically zero, leading to zero net rotation over a full cycle of shape deformations, the instantaneous rotations $\beta$ of the fish body couple displacements in the $x$- and $y$-directions, as evident from~\eqref{eq:eom_zeromom_trans}. 
For the cyclic shape changes in Fig.~\ref{fig:fishRace}(A), the amplitude of $y$-excursions is small but so is the net displacement in the $x$-direction; meanwhile 
in Fig.~\ref{fig:fishRace}(C), both the net displacement in the $x$-direction and the amplitude of the $y$-excursions are large. The fastest fish is the one that maximizes forward motion while minimizing lateral movements, as shown in Fig.~\ref{fig:fishRace}(B) and recapitulated in the RL result shown in Fig.~\ref{fig:fishRace}(D). It is worth emphasizing that the RL policy arrives at this optimal solution merely by sampling observations, actions, and rewards, with no prior or developed knowledge of the physics of the problem. 

Next, we investigated the effect of the initial orientation $\beta(0)\in [-\pi,\pi]$ on the amount of control effort required to turn and swim parallel to the $x$-axis. Fig.~\ref{fig:driftless}(A) shows three examples of fish following the trained policy starting from three distinct initial orientations $\beta(0) = 0,\pi/3$, and $2\pi/3$ and a straight configuration centered at the origin of the physical space. 
To measure the actuation effort needed in these motions, we used the integral $\int^{\tau}_{0}T_{\rm shape}\mathrm d t$ of the actuation energy
$T_{\rm shape} = \frac{1}{2}(J+m_{2}a^{2})(\dot{\alpha}^{2}_{1}+\dot{\alpha}^{2}_{2}),$
which is the energy imparted to the fluid by the fish shape changes; see Appendix A.  
Fig.~\ref{fig:driftless}(B) shows  the actuation effort versus the initial orientation of a straight fish. Here the fish was instructed by the stochastic policy with the same action noise as in the training process. The actuation effort, as well as its variation due to noise, is larger for larger turning angles.

Lastly, we examined the behavior and effort of a fish swimming instructed to reach a known target in an environment with zero drift.
Fig.~\ref{fig:driftless}(C) shows examples of the fish swimming motion in the physical space for targets located at $\psi = \pi/6$ and $\psi = 5\pi/6$. All tests ran until the fish reaches the target or a maximum interval of 1000 time steps is exceeded. We varied the target angular position while maintaining the fish initial shape and orientation (the fish always started straight in the $x$-direction), and we calculated the actuation effort as a function of the target orientation; see Fig.~\ref{fig:driftless}(D). The actuation effort increases as the target moves from the front to the back of the fish, because it requires larger turns in order for the fish to align its heading direction with the direction of the target; this is consistent with our findings based on the direction-control policy. It is worth emphasizing that the direction-control task is equivalent to the target-seeking task with the target placed at $x=+\infty$. it is also worth noting that the reward function is measured from the nose of the fish, thus breaking the fore-aft symmetry of the fish, and leading to control action that favor turning and swimming head first towards the target.

We tested the behavior and effort of the target-seeking fish in the presence of non-zero drift. 
Fig.~\ref{fig:target}(A) and (C) show a comparison between the naive and drift-aware policies for targets located at $\psi = \pi/6$ and $\psi = 5\pi/6$, with drift of magnitude of 0.1 pointing to the $x$-direction and the $-y$-direction, respectively. The naive policy (grey lines) is able to reach the target, even though it does not directly observe the drift, albeit following different actions and swimming trajectories than the drift-aware policy (orange lines). Specifically, when following the drift-aware policy, the fish curls less when the drift is helpful and curls more when the drift is unfavorable. We assessed the performance of the two policies for various drift magnitudes and directions. In Fig.~\ref{fig:target}(B), we calculated the actuation effort as a function of the drift magnitude with a fixed drift direction for two target locations.  The naive policy outperforms the drift-aware policy for small drift (drift magnitudes less than 0.025) even when the drift is for adversarial, but the naive policy loses or even fails to finish the task when the drift is large, especially when the drift is in the adverse direction to the target location. This implies that it might be wise to discard some sensory input (observations) when the perturbation due to drift is weak, especially if these extra observations act more like a distraction than a clue.  But as the perturbation gets stronger, it is necessary to take more observations into account. Both the naive and drift-aware policies are not able to complete the task in the given amount of time when the drift magnitude is very large and its direction is adversarial to the target location. This is because the shape actuation has no direct control over the drift itself.
In Fig.~\ref{fig:target}(D), we fixed the magnitude of drift to 0.15 and changed its direction. Using the actuation effort as our performance metric as before, the drift-aware policy has better or similar performance under all tested conditions. The naive policy fails when the drift acts in the adverse direction relative to the target while the drift-aware policy is always able to reach the target before the episode terminates.

\section{Concluding Remarks}
\label{sec:discussion}

We considered a three-link fish swimming in potential flow. We reviewed that swimming in potential flow can be expressed as a combination of a dynamic phase (drift) and geometric phase (driftless) over the shape of fish body deformation~\cite{Kanso2005a}. In the driftless case, net locomotion is purely determined by the fish shape deformations, and  geometric techniques can be used for gait design and motion planning over the fish shape space~\cite{Melli2006,Hatton2010}, but shape actuation cannot control the drift itself. Yet, even in the driftless regime, motion planning starting from arbitrary fish orientation and shape is not trivial. In this paper, we applied model-free reinforcement learning techniques for controlling the fish motion, and we arrived at optimal policies for swimming (i) in a desired  direction, and (ii) towards a target in the absence and presence of drift. The RL based policies produce behavior that is robust to variations in the fish initial shape and orientation and target location. We used the actuation effort as a measure of the policy performance under various initial conditions and in various environments, and we quantified the robustness of the RL policies to the presence of drift. We found that although the fish has no control over the drift itself, the fish learns to take advantage of the presence of moderate drift to reach its target. Therefore, it might be wise to discard some sensory input (observations) for weak signals because these extra observations could act more like a distraction than a clue. We also found that for small drift, the drift-naive policy outperforms the drift-aware policy even when drift is adversarial to the target location. However, large adversarial drift hinders the fish ability to locate the target. Importantly, these insights into the RL policies were achievable by combining tools from geometric mechanics with RL-based control.  Geometric tools such as the concept of shape space, combined with the RL notion of the action space, provided a useful and novel framework for visualizing and interpreting the RL policies as action vector fields over the shape space. 

A few comments on the advantages and limitations of our implementation are in order. 
Despite algorithmic advantages, obtaining an RL policy is computationally costly, especially when the environment simulator involves high-fidelity fluid-structure interaction models. To circumvent this problem, recent work on training fish to swim uses a limited set of observations and actions \cite{Verma2018}. For example, the zebrafish model of \cite{Verma2018} allows only 5 discrete actions, each corresponding to a fixed amplitude of body curvature change. 
Reduced order fluid models, such as the potential flow model used here, offer an enticing framework for designing control laws that can later be tested and refined using more realistic flow environments, as done in \cite{Eldredge2006} for a manually-designed swimming gait in \cite{Kanso2005a}.
Specifically, in the simplified potential flow environment employed here, we are able to train continuous action policies using a rich set of observations, with an eye on probing the performance of these policies in more realistic flow environments in future work.


\section*{Acknowledgments}
Kanso would like to acknowledge support from the Office of Naval Research through the grants N00014-17-1-2287 and N00014-17-1-2062. The authors are thankful for useful discussions with Dr. Yuval Tassa.

\bibliographystyle{vancouver}
\bibliography{references}

\appendix
\counterwithin{figure}{section}
\counterwithin{algorithm}{section}


\section{Physics of the fish model} We review the derivation of the equations of motion governing the swimming of an articulated three-link fish in potential flow (Fig.~\ref{fig:fishRL}).

\label{app:fishmath}
\subsection{Fish kinematics}
\label{app:addedmass}
Consider planar motions of the three-link fish. Let $\mathbf x = (x, y)$ denote the position of the center of mass $G$ of the middle link, and let $\beta$ denote the orientation of the fish relative to a fixed inertial frame, here taken to be the angle between the $x$-axis and the major axis of symmetry of the middle link. Let $\alpha_1$ and $\alpha_2$ be the rotation angles of the front link relative to the middle link  and the middle link relative to the rear link, that is to say, $(\alpha_1,\alpha_2)$ represents the shape of the three-link fish. It is convenient for the following development to introduce a body-fixed frame $(\mathbf b_{1},\mathbf b_{2},\mathbf b_{3})$, attached at $G$ and co-rotating with the middle link. This body-fixed frame is related to the inertial frame $(\mathbf e_{1},\mathbf e_{2},\mathbf e_{3})$ via a rigid-body rotation such that 
$\mathbf e_{1} = \cos\beta \mathbf b_{1} - \sin \beta \mathbf b_{2}$, $\mathbf e_{2} = \sin\beta \mathbf b_{1} + \cos \beta \mathbf b_{2}$, and $\mathbf e_{3} = \mathbf b_{3}$.


The velocity $(\dot{x}, \dot{y})$ of the center of mass of the middle link, when expressed in the body-fixed frame, is given by 
\begin{equation}
\mathbf{v} = v_1\mathbf b_{1} +v_2 \mathbf b_{2} 
=  ( \dot{x} \cos\beta + \dot{y}\sin\beta )\mathbf b_{1} 
+ ( -\dot{x} \sin\beta + \dot{y}\cos\beta)\mathbf b_{2}
\label{eq:velapp}
\end{equation}
Assuming all three links are made of identical ellipsoids of half-length $a$, half-width $b$, and half-height $c$, the velocities of the centers of mass $G_1$ and $G_2$ of the front and rear link, expressed in the body-fixed frame of the middle link, are given by ($i = 1,2$, denote the front and rear link, respectively)
\begin{equation}
\mathbf{v}_i = (v_1 \mp a \dot\alpha_i\sin \alpha_i - a\dot{\beta}\sin\alpha_{i})\mathbf b_{1}
+ ( v_2 \pm a\dot{\beta} + a \dot\alpha_{i}\cos\alpha_{i} \pm a\dot{\beta}\cos\alpha_{i})\mathbf b_{2}.
\end{equation}
The angular velocities of the middle, front, and rear links are given by $\dot{\beta}$, $\dot{\beta} + \dot{\alpha}_1$, and $\dot{\beta} - \dot{\alpha}_2$ respectively. For our simulations, we used $a:b:c = 5:1:5$ as the geometry of the ellipsoids.
\subsection{Kinetic energy of the articulated body} In the absence of the fluid, the kinetic energy of the articulated three-link body is given by
\begin{equation}
\label{eq:Ts}
T_{\rm body} =  \dfrac{1}{2} m_s \mathbf{v}\cdot\mathbf{v} + \dfrac{1}{2} J_s \dot{\beta}^2 +  
 \dfrac{1}{2} \sum_{i} \left[m_s \mathbf{v}_i\cdot \mathbf{v}_i + J_s (\dot{\beta} \pm \dot{\alpha}_{i})^2 \right]
\end{equation}
where  $m_s = \frac{4}{3}\pi abc\rho_{s}$ and $J_s = \frac{1}{5}(a^{2}+b^{2})m_{s}$ are the mass and moment of inertia of each solid link with $\rho_{s}$ the density of the links.

\subsection{Kinetic energy of the fluid} The three-link fish is submerged in an unbounded domain of incompressible and irrotational fluid, such that the fluid velocity $\mathbf{u} = \nabla \phi$ can be expressed as the gradient of a potential function $\phi$. It is a standard result in potential flow theory that the kinetic energy of the fluid can be expressed in terms of the variables of the submerged solid~\cite{Lamb1945, Kanso2005a,Kanso2005b}. In the case of a single ellipsoid, the kinetic energy of the fluid is given by $T_{\rm fluid} = [( m_{1a} v_1^2 + m_{2a} v_2^2) +  J_a \dot{\beta}^2]/2$, where $m_{1a}$, $m_{2a}$ and $J_a$ are the added mass and added moment of inertia due to the presence of the fluid, expressed in a body-fixed frame that coincides with the major and minor axes of the ellipsoid. These quantities depend on the geometric properties $a, b, c$ of the submerged ellipsoid, as are given in the Appendix B of~\cite{Leonard1997}. For a non-spherical body, the added masses $m_{1a}$, $m_{2a}$ depend on the direction of motion: the added mass is larger when moving in the direction of the minor axis of symmetry of the ellipsoid, that is to say, in the transverse direction, hence $m_{1a} \leq m_{2a}$.

In the case of the three-link fish, the kinetic energy of the fluid is of the form
\begin{equation}
\begin{split}
\label{eq:Tf}
 T_{\rm fluid} = &  \frac{1}{2}m_{1a}v_1^{2} + \frac{1}{2}m_{2a}v_2^{2} + \frac{1}{2}J_a\dot\beta^{2}
 + \frac{1}{2}\sum_{i}J_a\left(\dot\beta \pm \dot\alpha_{i}\right)^{2} \\
& \quad +\frac{1}{2}\sum_{i}m_{1a}\left(v_1 \cos{\alpha_{i}} \pm v_2\sin{\alpha_{i}}+a\dot\beta\sin\alpha_{i}\right)^{2}\\
& \quad + \frac{1}{2}\sum_{i}m_{2a}\left(\mp v_1 \sin{\alpha_{i}} + v_2\cos{\alpha_{i}}\pm a\dot\beta\cos\alpha_{i}\pm a\dot\beta + a\dot\alpha_{i}\right)^{2}.
\end{split}
\end{equation} 
Here we transform the velocity components of head and tail by $\alpha_{1}$ and $\alpha_{2}$, respectively, to match the added mass components.

\subsection{Kinetic energy of the body-fluid system} The kinetic energy of the fish-fluid system is obtained by taking the sum of Eq.~\ref{eq:Ts} and Eq.~\ref{eq:Tf}, which can be expressed in matrix form as follows:
\begin{equation}
\begin{split}
T &= T_{\rm body} + T_{\rm fluid} 
=\frac{1}{2}
\left[
\begin{array}{c}
v_1\\
v_2\\[1ex]
\dot\beta\\
\dot\alpha_1\\
\dot\alpha_2\\
\end{array}
\right]^{\mathsf T}
\left[
\begin{array}{ccc|cc}
 & & & & \\
 &\mathbb I_{\rm lock}& & & \mathbb I_{\rm couple}\\
&&&&\\\cline{1-5}
&&&&\\
&\mathbb I^{\mathsf T}_{\rm couple}& & & \mathbb I_{\rm shape}\\
\end{array}
\right]
\left[
\begin{array}{c}
v_1\\
v_2\\[1ex]
\dot\beta\\
\dot\alpha_{1}\\
\dot\alpha_{2}
\end{array}
\right],
\end{split}
\end{equation}
Here, $\mathbb I_{\rm lock}$ is a $3\times 3$ locked mass matrix, function of $\alpha_{1}$ and $\alpha_{2}$, 
\begin{equation}
\label{eq:Ilock}
\mathbb I_{\rm lock} =
\left[
\begin{array}{ccc}
\mathbb M && \mathbb H\\
&&\\[-0.5em]
\mathbb H^{\mathsf T}& & \mathbb J
\end{array}
\right],
\end{equation}
where $\mathbb M$ is a $2\times 2$ mass matrix given by
\begin{equation}
\label{eq:M}
\mathbb M =
\left[
\begin{array}{c c}
m_{1} \left(1+ \sum_{i}\cos^{2}\alpha_{i}\right) + m_{2}\sum_{i}\sin^{2}\alpha_{i} & \frac{1}{2}(m_{1}-m_{2})(\sin 2\alpha_{1}-\sin 2\alpha_{2})\\
 \frac{1}{2}(m_{1}-m_{2})(\sin 2\alpha_{1}-\sin 2\alpha_{2}) &m_{2} \left(1+ \sum_{i}\cos^{2}\alpha_{i}\right) + m_{1}\sum_{i}\sin^{2}\alpha_{i}
\end{array}
\right],
\end{equation}
$\mathbb J$ is a moment-of-inertia scalar given by
\begin{equation}
    \label{eq:J}
    \mathbb J = 3J+m_{1}a^{2}\sum_{i}\sin^{2}\alpha_{i} + m_{2}a^{2}\sum_{i}\left(1 + \cos\alpha_{i}\right)^{2},
\end{equation}
and $\mathbb{H}$ is given by
\begin{equation}
\label{eq:H}
\mathbb H =
\left[
\begin{array}{c}
\frac{1}{2}(m_{1}-m_{2})a\sum_{i}\sin 2\alpha_{i}- m_{2} a\sum_{i}\sin\alpha_{i}\\
\frac{1}{2}(m_{1}-m_{2})a(\cos 2\alpha_{2}-\cos 2\alpha_{1}) + m_{2}a(\cos\alpha_{1}-\cos\alpha_{2})
\end{array}
\right].
\end{equation}
Here we used $m_{1} = m_{s}+m_{1a}$, $m_{2} = m_{s}+m_{2a}$, and $J = J_{s}+J_a$. Note that $\mathbb{H}$ couples the translational and rotational motion of the articulated body. In the case of single ellipsoid, $\mathbb{H}$ is identically zero.

Further, $\mathbb I_{\rm couple}$ is a $3\times 2$ matrix that couples rigid body motion with shape deformation,
\begin{equation}
\label{eq:Icouple}
\mathbb I_{\rm couple} =
\left[
\begin{array}{c c}
-m_{2}a\sin\alpha_{1} & m_{2}a\sin\alpha_{2} \\
m_{2}a\cos\alpha_{1} & m_{2}a\cos\alpha_{2}\\
J+m_{2}a^{2}(1+\cos\alpha_{1}) & -J-m_{2}a^{2}(1+\cos\alpha_{2})
\end{array}
\right].
\end{equation}
Finally, $\mathbb I_{\rm shape}$ is a $2\times 2$ matrix associated with shape deformation,
\begin{equation}
\mathbb I_{\rm shape} =
\left[
\begin{array}{c c}
J+m_{2}a^{2} & 0 \\
0 & J+m_{2}a^{2}
\end{array}
\right].
\end{equation}
The total linear and angular momenta $P_1$, $P_2$, and $\Pi$ expressed in body frame are given by
$P_{1} = \partial T/ \partial v_1, P_{2} = \partial T/\partial v_2, \Pi = \partial T/\partial \dot\beta$ to arrive at~\eqref{eq:totalmomenta} of the main text
\begin{equation}
\left[\begin{array}{c}
P_{1} \\ P_{2} \\ \Pi
\end{array}\right] = \mathbb I_{\rm lock}
\left[\begin{array}{c}
v_1 \\ v_2 \\[1ex] \dot\beta
\end{array}\right] + \mathbb I_{\rm couple}
\left[\begin{array}{c}
\dot\alpha_{1} \\ \dot\alpha_{2} 
\end{array}\right].
\end{equation}



\section{Proximal Policy Optimization (PPO) Algorithms}
\label{app:ppo}

\begin{figure}[t]
\begin{algorithm}[H]
\caption{Environment Simulation}\label{alg:sim}
\begin{algorithmic}[1]

\For{time step $t = 0,1,...$ }

\If{$t=0$ \textbf{or} episode terminates} 
 \State store time step of episode termination,
 \State reset state $s_t\sim P(s_0)$
 \State evaluate observation: $o_t \sim o(s_{t})$
\EndIf{}
\State sample action from policy $a_t \sim \pi_{\theta}(a_t|o_t)$ 
 \State evolve next state according to fish physics $s_{t+1} \sim P(s_{t+1} |s_t , a_t)$ 
 \State evaluate next observation $o_{t+1} \sim o(s_{t+1})$ and reward $r_t \sim r(a_{t},o_{t+1})$
 \If{$t=0$ \textbf{or} mod$(t,N)\neq0$}
 \State append current action, observation, reward, and probability of sampling the action to assemble vectors \Statex $a_{N\times n_a}, o_{N\times n_o}, r_{N \times 1}$, and $\pi_{\theta_{\textrm{old}}}(a|o)_{N\times 1}$
 \Else
 \State update agent networks according to Algorithm~\ref{alg:update}
 \EndIf
\EndFor

\end{algorithmic}
\end{algorithm}
\end{figure}

\begin{figure}[t]
\begin{algorithm}[H]
\caption{Updating the Agent}\label{alg:update}
\begin{algorithmic}[1]
\For{update epoch number $\kappa = 0,1, \hdots K$}
\State compute the truncated return using rewards $r_{N\times 1}$ and assemble into vector $R_{N \times 1}$
\State estimate infinite-horizon return using $R_{N \times 1}$ and $V_T=V_\phi(o_T)$ if bootstrapping is desired (see Eq.\ref{eq:hatQK})
\State using $o_{N\times n_o}$ and value function $V_{\phi}$, evaluate expected returns at each time step and store into $V_{N \times 1}$
\State compute the advantage $A = R_{N \times 1} - V_{N \times 1}$ and normalize by its mean and variance if desired
\State evaluate the probability of realizing $a_{N\times n_a}$ based on $o_{N\times n_o}$ for the policy $\pi_\theta$, and store to $\pi_{\theta}(a|o)_{N\times 1}$
\vspace*{0.25em}
\State compute the action-likelihood ratio: $\varrho_\theta= \dfrac{\pi_{\theta}(a|o)_{N\times 1}}{\pi_{\theta_{\textrm{old}}} (a|o)_{N\times 1}}$
\vspace*{0.25em}
\State compute clipped surrogate loss function: 
$\mathcal{L}_{\text{clip}}(\theta) = \operatorname{mean}\left[\min\left[\varrho_\theta\cdot A, \textrm{clip} (\varrho_\theta, 1-\epsilon,1+\epsilon)\cdot A\right]\right]$
\vspace*{0.25em}
\State compute the value-function loss: $ \mathcal{L}_{\textrm{value}}(\phi) = 0.5\cdot\operatorname{mean}\left[(R_{N \times 1} - V_{N \times 1})^2\right]$
\vspace*{0.25em}
\State compute the total loss: $\mathcal{L}(\theta, \phi) = -\mathcal{L}_{\textrm{clip}} (\theta)  + \mathcal{L}_{\textrm{value}} (\phi) - \alpha\cdot \operatorname{entropy}\left[\pi_{\theta}\right]$
\vspace*{0.25em}
\State update parameters $(\theta, \phi)$ to minimize the total loss using a gradient based optimizer (\textit{e.g.}, Adam \cite{Kingma2014})
\EndFor
\end{algorithmic}
\end{algorithm}
\end{figure}

We implement the clipped advantage Proximal Policy Optimization (PPO) method proposed by \cite{Schulman2017b} for our RL training. PPO maximizes a surrogate objective that clips off unwanted changes when the policy deviates too much from the policy of the previous cycle to ensure faster and more robust convergence. We refer readers to the original reference cited above as well as the \href{https://spinningup.openai.com/en/latest/algorithms/ppo.html}{OpenAI's documentation} of the PPO algorithm and their \href{https://github.com/openai/baselines/tree/master/baselines/ppo2}{baseline implementations} for a thorough explanation of the theory and details behind this method.

Our implementation can be separated into two parts. The main loop simulates the environment using action sequence $a_t$ generated by the agent, and stores the observed rollouts for future updates; see Algorithm~\ref{alg:sim}. Note that $n_o$ and $n_a$ are used to indicate the number of observable states and actions. Equations describing the fish-fluid interactions were integrated numerically using adaptive time step, explicit RK45 method between each decision step of $0.1$ unit of time. This choice of decision time step-size limits the maximum rotation allowed for the fish head and tail to be $0.1$ radian per step.

Parameters of the actor-critic networks of the RL agent are updated every $N$ time steps for $K$ epochs. Here the value of $K$ is chosen to be 80 and the value of $N$ is set to 4050, an integer multiple of the episode length 150. For simplicity, we assume our continuous action variables follows a multivariate normally-distributed policy $\pi_{\theta}$ with mean value represented by a neural network parameterized by $\theta$ and constant diagonal covariance matrices, and the critic / value function $V_{\phi}(o_t)$ is also represented by a neural network with parameters $\phi$.
Specifically, both the mean policy and value function are implemented as feed-forward neural network with two hidden layers and $\tanh$ activation functions. The sizes of the two hidden layers were fixed to 64 and 32, respectively. Each diagonal entry of the covariance matrix is set to $0.5^{2}$.
Finally, using the collected trajectories during the previous $N$ time steps, the parameters $\theta,\phi$ are updated according a total loss function $\mathcal{L}(\theta, \phi)$ via a back-propagating gradient based optimizer; see Algorithm~\ref{alg:update}. Note that since we did not perform systematic hyper-parameter tuning, readers might want to explore different values for better performance.

Another important side-note is that since it is in general impossible to obtain unrealized infinite horizon return $R_t=\Sigma_{t'}^\infty \gamma^{t'-t}r_{t'}$, we need to choose an appropriate estimator of this value based on finite length simulations. We can either simply truncate rewards after some step $k$ by using
\begin{align}
    \left. \hat{R}_t\right|_{\text{truncation}} &= r_t + \gamma r_{t+1} + \gamma^2 r_{t+2} + \dots + \gamma^{k-1} r_{t+k}, \label{eq:hatQ_RtK} 
\end{align}
or we can use the trained value function (critic) to approximate the residual contribution to the return via $k$-step bootstrapping
\begin{equation}
    \left. \hat{R}_t\right|_{\text{bootstrapping}} = r_t + \gamma r_{t+1} + \gamma^2 r_{t+2} + \dots + \gamma^{k} V_\phi(o_{t+k+1}). \label{eq:hatQK} 
\end{equation}
\begin{figure*}[!t]
\centering
\includegraphics[scale=1]{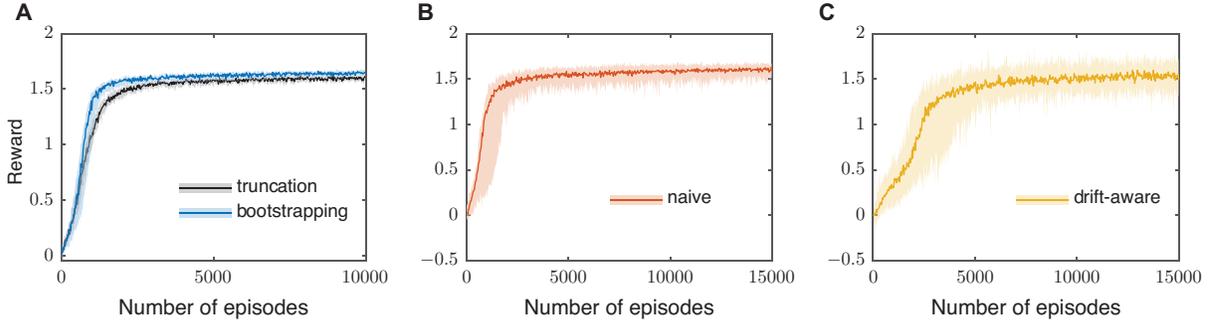}
\caption{\textbf{Evolution of rewards during the training process.} \textsf{A}. Total rewards per episode achieved by policies trained to swim parallel to the $x$-axis in a driftless environment using bootstrapped (blue) and truncated (black) return estimates. Here solid lines indicate the median, and the shaded region shows the variation between 25\--75 percentile for 24 runs of the learning algorithm. \textsf{B}. and \textsf{C}. Total rewards obtained by policies trained to swim towards a given target, both of which adopt bootstrapped return estimates. Red in \textsf{B} reresents naive policies trained in driftless environment, while yellow in \textsf{C} represents policies trained in the presence of drift, with drift magnitude and direction supplied as additional observations to the policy. Again, lines and shaded regions indicate median and 25\--75 percentile range respectively.}
\label{fig:rewVsEpi}
\end{figure*}
We compared these two approaches for the direction-based task and observed that bootstrapping results in faster convergence and higher rewards in general; see Fig.~\ref{fig:rewVsEpi}(A). As a result, bootstrapping is used for all tasks depicted in the main text, where $k$ was determined by the number of available future rewards. Namely, $k$ decreases from 149 to 0 as the number of time steps increased from 1 to 150 in each episode.
In addition, We show the difference in training rewards and convergence speed between the naive policy and the drift-aware policy in Fig.~\ref{fig:rewVsEpi}(B). In general, the inclusion of more observations increased the time to convergence and variance in training rewards.

Lastly, we invite interested readers to visit our source repository at \url{https://github.com/mjysh/RL3linkFish} for the complete details of our implementation.

\end{document}